\documentclass[aps,prx,twocolumn,superscriptaddress,10pt]{revtex4-2}
\usepackage{amsmath,amssymb,graphicx,bm}
\usepackage{float}

\usepackage{hyperref}
\hypersetup{
    colorlinks=true,       
    linkcolor=blue,        
    citecolor=blue,        
    urlcolor=blue,         
    breaklinks=true        
}

\begin{document}

\title{Microsecond-Programmable Synthetic Quantum Matter: \\ Circumventing the Adiabatic Bottleneck for Heisenberg-Limited Metrology}

\author{Soi-Chan Lei}
\affiliation{Independent Researcher, Macau 999078, China}

\begin{abstract}
Generating macroscopic quantum entanglement is central to quantum metrology. While traversing a quantum critical point can yield Heisenberg-limited precision, adiabatic preparation schemes are typically restricted by slow dynamics and environmental decoherence. Here, we propose a non-adiabatic Floquet protocol in a single-cavity Bose-Einstein condensate to circumvent these limitations. By applying a sequence of rapid $\pi/2$ rotations and phase jumps, we dynamically synthesize effective Two-Axis Twisting (TAT) interactions from native longitudinal cavity couplings. To evaluate this protocol beyond the mean-field limit, we develop an exact finite-state automaton Matrix Product Operator (MPO) encoding. This formulation strictly bounds the MPO virtual dimension to $\mathcal{O}(1)$, enabling exact Density Matrix Renormalization Group (DMRG) and Time-Dependent Variational Principle (TDVP) simulations for arbitrarily large systems. While static evaluations capture a finite-size quantum shift of the critical boundary and a logarithmic divergence of entanglement entropy, we apply this method to realistically large ensembles ($N=10^3$) to rigorously track the non-equilibrium Floquet TAT dynamics and the inevitable onset of environmental decoherence. We demonstrate that the dynamic Floquet quench rapidly generates Heisenberg-limited spin squeezing ($\xi_R^2 \sim 1/N$). Furthermore, the microsecond timescale of this dynamic protocol effectively decouples the generation process from typical millisecond experimental decoherence. Phenomenological dissipative models reveal a shifted optimal squeezing time near $t \approx 3.26~\mu\mathrm{s}$. In this regime, the exact evolution exhibits rapid entanglement growth that fundamentally restricts efficient classical simulability. By providing hardware-compatible pulse sequences, this work establishes a rigorous and feasible route for realizing Heisenberg-limited metrology in current cavity-QED platforms.
\end{abstract} 

\maketitle

\section{Introduction}

Synthetic quantum matter provides a highly controllable platform for advancing quantum metrology, many-body physics, and information processing \cite{georgescu2014quantum, pezze2018quantum}. A key focus in this field is the engineering of collective spin models, such as the XXZ model, where the interplay between longitudinal density-density interactions and transverse flip-flop exchanges drives macroscopic quantum phase transitions. 

While collective models have been realized in platforms such as dipolar quantum gases and Rydberg atom arrays \cite{lahaye2009thephysics, browaeys2020many}, they are often limited by fixed interaction geometries and finite tuning rates. Cavity QED offers a versatile alternative: exploiting all-optical Raman transitions and far-detuned AC Stark shifts enables the simultaneous synthesis of independent, tunable, all-to-all transverse and longitudinal interactions \cite{periwal2021programmable, norcia2018cavity}. 

In this work, we utilize this optical tunability to construct a macroscopic XXZ spin model in a single-cavity Bose-Einstein condensate (BEC). Our framework achieves microsecond-scale independent control over both interaction channels, dynamically decoupling the transverse exchange ($\lambda$) from the longitudinal density coupling ($\chi$). Furthermore, applying Floquet engineering \cite{else2016floquet} allows us to extend the dynamical regime of this platform. While the static Hamiltonian generates One-Axis Twisting (OAT) dynamics, we demonstrate that a specific Floquet sequence of rapid ($\pi/2$) rotations and ($\pi$) phase jumps dynamically transforms the native longitudinal interaction into a Two-Axis Twisting (TAT) Hamiltonian.

This Floquet approach addresses a primary challenge in quantum metrology. Although adiabatic preparation near a quantum critical point theoretically yields Heisenberg-limited precision ($F_Q \sim N^2$) \cite{rams2018quantum}, critical slowing down makes adiabatic protocols highly susceptible to realistic decoherence. We bypass this limitation by synthesizing TAT through a fast, non-adiabatic quench, driving the initial coherent state to Heisenberg-limited spin squeezing ($\xi_R^2 \sim \mathcal{O}(1/N)$) purely via non-equilibrium dynamics.

To evaluate this protocol beyond mean-field approximations, we construct an exact finite-state automaton MPO encoding (see Appendix~\ref{app:mpo_automaton}). This mathematical construction analytically bounds the virtual bond dimension strictly to $\mathcal{O}(1)$, enabling exact DMRG and TDVP simulations for arbitrarily large systems. Showcasing this capability at a realistic experimental scale of $N=10^4$ atoms, our exact numerical analysis captures a finite-size quantum shift of the critical point and a logarithmic divergence of entanglement entropy, effects that fundamentally elude classical mean-field theories. Crucially, to transcend the metrological limitations of these static critical phenomena, we subsequently deploy the engineered TAT Hamiltonian as a specialized dynamic protocol. This allows us to rigorously track the non-equilibrium Floquet TAT dynamics and the inevitable onset of environmental decoherence for realistically large ensembles ($N=10^3$). It is this exact real-time tracking that definitively establishes the robust, dynamically generated Heisenberg-limited advantage.

This framework contributes to several areas of quantum many-body physics and metrology:

\textbf{1. Exact Simulation of Macroscopic Entanglement:} The $\mathcal{O}(1)$ MPO construction provides a scalable numerical method to benchmark macroscopic many-body dynamics, allowing exact calculation of highly entangled states beyond mean-field limitations.

\textbf{2. Bypassing the Adiabatic Bottleneck:} By synthesizing TAT through Floquet modulation, we demonstrate that Heisenberg-limited spin squeezing can be achieved via fast non-equilibrium quenches, avoiding the slow adiabatic preparation barrier and the associated Kibble
Zurek excitations \cite{macri2016loschmidt}.

\textbf{3. Mitigation of Environmental Decoherence:} Operating on a microsecond ($\mu$s) timescale, the optical protocol is significantly faster than typical millisecond (ms) decoherence channels in high-finesse cavities, allowing metrological states to be extracted before dissipation dominates.

\textbf{4. Connections to Unified Many-Body Models:} The dynamics of the engineered XXZ model map to several canonical systems, connecting the optical implementation to the Richardson-BCS model in superconductivity, the Lipkin-Meshkov-Glick (LMG) model in nuclear physics, and collective Dicke superradiance \cite{richardson1963restricted, lipkin1965validity, dicke1954coherence}.

By combining an exact $\mathcal{O}(1)$ numerical framework with hardware-compatible Floquet pulse sequences, our results provide a complete theoretical and operational methodology for realizing Heisenberg-limited metrology in synthetic quantum platforms.

\section{The Engineered Collective XXZ Hamiltonian}
\label{sec.II}
To establish the theoretical framework for the synthetic quantum matter, we construct a microscopic Hamiltonian that describes the engineered cavity-mediated interactions. The effective dynamics are governed by the interplay and tuning ratio between the longitudinal ($\chi$) and transverse ($\lambda$) interactions.

\subsection{Physical Origins: Orthogonal Optical Dual-Driving}

The Hamiltonian is synthesized using two independent optical mechanisms, ensuring decoupled control over the interaction channels:

\textbf{Longitudinal Interaction ($\chi$):} The longitudinal density-density interaction is generated via a far-detuned cavity-mediated AC Stark shift \cite{brennecke2007cavity}. By dispersively coupling the cavity field to the collective spin ($H_{\mathrm{disp}} \propto a^\dagger a J_z$) and adiabatically eliminating the cavity mode \cite{hartmann2006strongly}, an effective nonlinear interaction is induced:
\begin{equation}
    H_{\parallel} = \frac{\chi}{N} \sum_{k \neq l}^N \sigma_k^z \sigma_l^z.
\end{equation}
A dual-frequency driving scheme orthogonal to the Raman transition isolates this longitudinal channel from transverse spin-exchange dynamics.

\textbf{Transverse Interaction ($\lambda$):} To synthesize the all-to-all transverse flip-flop exchange without interfering with $\chi$, we deploy a polarization-orthogonal driving scheme, typical in dual-color cavity QED. A distinct set of external control lasers generates the transverse term:
\begin{equation}
    H_{\perp} = \frac{\lambda}{2N} \sum_{k \neq l}^N \left( \sigma_k^x \sigma_l^x + \sigma_k^y \sigma_l^y \right).
\end{equation}
By utilizing orthogonal spatial modes and sufficient frequency separation, the transverse drive is strictly decoupled from the longitudinal AC Stark effect. To ensure a well-defined thermodynamic limit, we apply Kac scaling \cite{koffel2012entanglement}, normalizing both infinite-range interactions by the total particle number $N$.

\subsection{Microscopic Lattice and Macroscopic Collective Models}

Combining the bare atomic transition energy (parameterized by the effective detuning $\omega_0$) with the engineered interactions yields the exact microscopic Hamiltonian:
\begin{equation}
    H = \frac{\omega_0}{2} \sum_{k=1}^N \sigma_k^z + \frac{\lambda}{2N} \sum_{k \neq l}^N (\sigma_k^x \sigma_l^x + \sigma_k^y \sigma_l^y) + \frac{\chi}{N} \sum_{k \neq l}^N \sigma_k^z \sigma_l^z.
    \label{eq:micro_H}
\end{equation}

Projecting Eq.~(\ref{eq:micro_H}) onto the fully symmetric collective spin manifold rewrites the system as a collective XXZ spin model:
\begin{equation}
    H_{\mathrm{coll}} = \omega_0 J_z + \frac{2\lambda}{N} (J_x^2 + J_y^2) + \frac{4\chi}{N} J_z^2,
    \label{eq:macro_H}
\end{equation}
where constant $\mathcal{O}(1)$ energy shifts from atomic self-interactions have been dropped. 

In the maximally symmetric Dicke manifold, the conserved total collective spin length $J^2 = J_x^2 + J_y^2 + J_z^2 = \frac{N}{2}(\frac{N}{2}+1)$ yielding $J_x^2 + J_y^2 = J(J+1) - J_z^2$, which is substituted into Eq.~(\ref{eq:macro_H}), the system maps exactly onto an effective Lipkin-Meshkov-Glick (LMG) Hamiltonian:
\begin{equation}
    H_{\mathrm{eff}} = \omega_0 J_z + \frac{1}{N}(4\chi - 2\lambda) J_z^2 + \mathcal{O}(N).
    \label{eq:eff_H}
\end{equation}
The constant $\mathcal{O}(N)$ term ($\frac{\lambda N}{2} + \lambda$) represents a rigid global energy shift. 

Eq.~(\ref{eq:macro_H}) reflects the microscopic dual-driving optical architecture, while the reduced form in Eq.~(\ref{eq:eff_H}) isolates the fundamental competition driving the ground-state quantum phase transition (QPT). Specifically, the linear term $\omega_0 J_z$ acts as an effective magnetic field favoring spin polarization, whereas the engineered nonlinear term $\frac{1}{N}(4\chi - 2\lambda) J_z^2$ promotes quantum fluctuations and macroscopic entanglement. 

The ground-state phase diagram is determined by the ratio of the effective interaction strength $|4\chi - 2\lambda|$ to the detuning $\omega_0$. The static Hamiltonian naturally generates OAT dynamics. The independent optical control over $\chi$ and $\lambda$ provides the prerequisite handles for implementing the subsequent Floquet modulation to dynamically synthesize TAT.

\begin{figure}[t]
    \centering
    \includegraphics[width=0.48\textwidth]{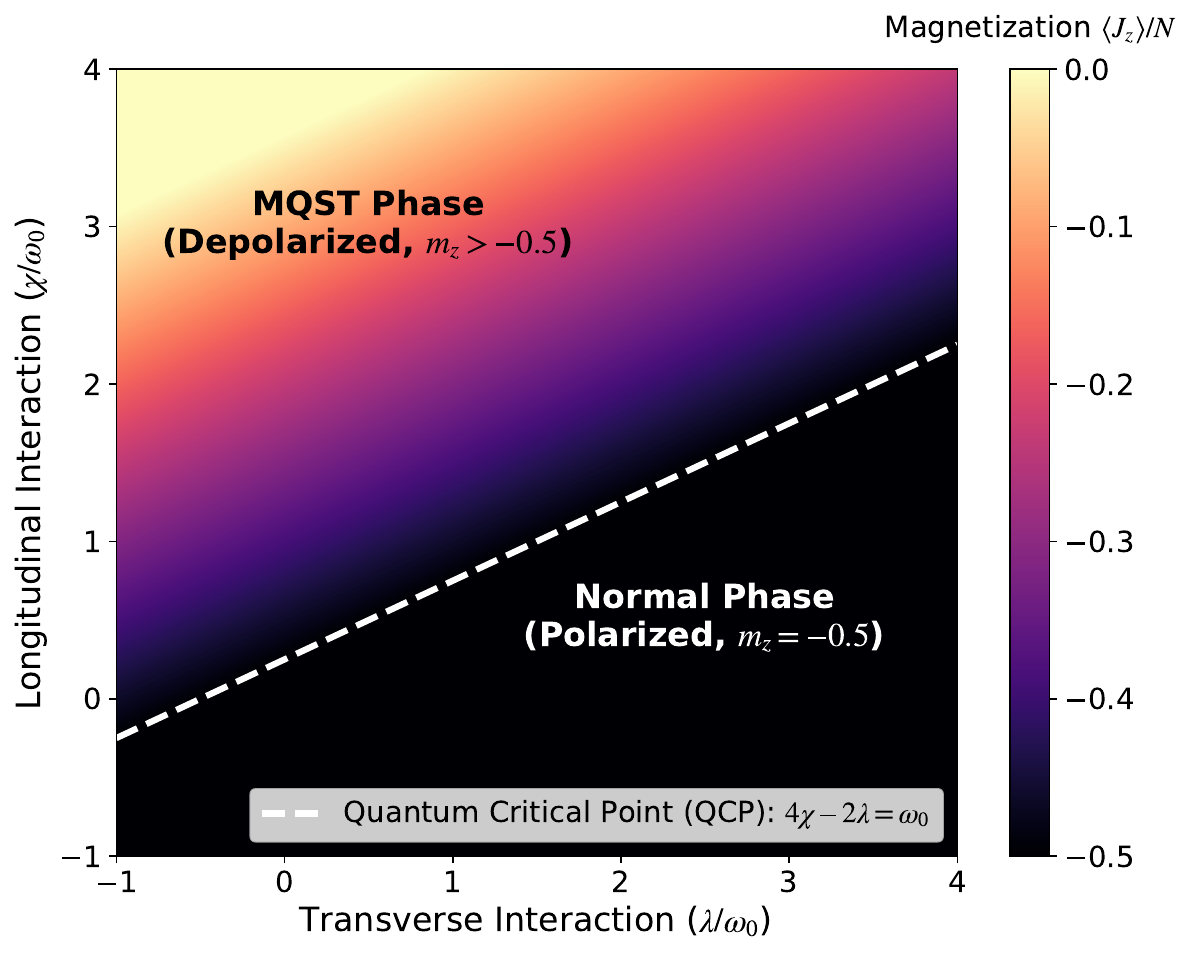}
    \caption{\textbf{Mean-Field Ground-State Phase Diagram of the Engineered XXZ Model.} The color map indicates the collective magnetization $m_z = \langle J_z \rangle / N$. The dashed white line denotes the semi-classical quantum critical point (QCP) boundary at $|4\chi - 2\lambda| = \omega_0$, separating the fully polarized Normal Phase from the depolarized Macroscopic Quantum Self-Trapping (MQST) Phase.}
    \label{fig:mf_phase_diagram}
\end{figure}

As illustrated in Fig.~\ref{fig:mf_phase_diagram}, mean-field theory locates the approximate critical boundaries in the $N \to \infty$ limit by assuming separable product states and explicitly neglecting quantum fluctuations. However, this semiclassical framework is fundamentally insufficient to capture the finite-size quantum shift or the metrological gain generated by the non-adiabatic Floquet TAT dynamics. Therefore, to rigorously evaluate the diverging non-local quantum correlations and extract the true macroscopic quantum advantage, we must employ the exact DMRG and TDVP algorithms introduced in Sec.~\ref{IV}.

Before the dynamical protocols, we note a key distinction. The full Hamiltonian (including $\omega_0 J_z$) is strictly retained for static DMRG calculations to faithfully map the QPT, while our metrological focus relies on non-equilibrium Floquet generation. All subsequent Floquet TAT formulations and real-time TDVP simulations are therefore performed in a frame rotating at the bare atomic transition frequency $\omega_0$. This cleanly eliminates the trivial free-evolution term $\omega_0 J_z$, strictly isolating and harnessing the nonlinear interaction dynamics.

\section{Quantum Metrology and the Limits of Spin Squeezing}
\label{sec.III}

To evaluate the metrological potential of the engineered XXZ model, we analyze its capacity to generate macroscopic quantum entanglement and surpass the Standard Quantum Limit (SQL). 

\subsection{Metrological Quantifiers and the Heisenberg Limit}

The metrological utility of a quantum state is quantified by the Wineland spin-squeezing parameter $\xi_R^2$ \cite{kitagawa1993squeezed}. A state surpasses the SQL when $\xi_R^2 < 1$. A more fundamental bound is established by the Quantum Cramér-Rao Bound (QCRB), $(\Delta\theta)^2 \ge 1/F_Q$, where $F_Q$ is the Quantum Fisher Information (QFI). For a maximally entangled state, the QFI scales as $F_Q \sim N^2$, defining the Heisenberg Limit (HL) \cite{pezze2018quantum}.

These two metrological quantifiers are related by the inequality $F_Q/N \ge 1/\xi_R^2$. Achieving Heisenberg-limited spin squeezing ($\xi_R^2 \sim \mathcal{O}(1/N)$) inherently guarantees a macroscopic enhancement of the QFI ($F_Q \sim \mathcal{O}(N^2)$). Given this rigorous mathematical bound, minimizing the experimentally accessible $\xi_R^2$ serves as a sufficient practical proxy for maximizing the QFI. The extraction of this quantum advantage thus translates to dynamically minimizing $\xi_R^2$ using the engineered Hamiltonian.

\subsection{Dynamical Entanglement Generation via Floquet Engineering}

We initialize the system in a coherent spin state (CSS) polarized along the $x$-axis. As shown in Sec.~\ref{sec.II}, the unmodulated static Hamiltonian trivially preserves $U(1)$ rotational symmetry within the collective manifold ($J_x^2 + J_y^2 = J^2 - J_z^2$). Consequently, the static competition between transverse and longitudinal terms merely renormalizes the interaction to an effective OAT form, $H_{\mathrm{eff}} \propto (4\chi - 2\lambda) J_z^2$. While OAT generates spin squeezing, the minimal squeezing parameter asymptotically scales only as $\xi_R^2 \sim N^{-2/3}$ \cite{kitagawa1993squeezed}. 

To achieve HL scaling ($\xi_R^2 \sim 1/N$), the system requires genuine TAT interactions, which continuously squeeze the uncertainty uniformly along one axis while anti-squeezing it along the orthogonal axis, thereby preventing detrimental Bloch-sphere wrap-around. To synthesize TAT dynamics, we employ a Floquet modulation scheme to deliberately break the inherent $U(1)$ symmetry that otherwise restricts the static system to OAT.

By utilizing external optical driving fields, we periodically pulse the system to induce global $\pi/2$ rotations. Specifically, applying a rapid $\pi/2$ pulse around the $x$-axis, given by $e^{-i(\pi/2)J_x}$, unitarily rotates the spin operators such that $e^{-i(\pi/2)J_x} J_z e^{i(\pi/2)J_x} = -J_y$, rigorously mapping the native longitudinal interaction $J_z^2$ into a transverse $J_y^2$ term. However, a naive Floquet scheme that simply alternates between $J_z^2$ and $J_y^2$ is physically insufficient, as the time-averaged effective interaction of such a sequence would be proportional to $(J_z^2 + J_y^2)$. Since $J_z^2 + J_y^2 = J^2 - J_x^2$, this combination trivially reduces back to a OAT form. To realize TAT dynamics, we must synthesize a crucial minus sign to invert the transverse interaction. Experimentally, this is achieved by inserting a global $\pi$ phase pulse between the two interaction blocks, cleanly converting the $J_y^2$ term into the required $-J_y^2$ term.

The resulting time-averaged Floquet Hamiltonian synthesizes the target symmetry-breaking interaction \cite{kitagawa1993squeezed, kajtoch2020squeezing}:
\begin{equation}
    H_{\mathrm{TAT}}^{\mathrm{eff}} \simeq \frac{\chi_{\mathrm{eff}}}{N} (J_z^2 - J_y^2).
    \label{eq:tat}
\end{equation}
Here, the approximation ($\simeq$) signifies the truncation of the Magnus expansion to its leading order, which is formally valid in the fast-pulsing limit (i.e., when the Floquet cycle is much shorter than the interaction timescale). 

According to average Hamiltonian theory (AHT), a symmetric driving sequence---evolving under $J_z^2$ and $-J_y^2$ for strictly equal time intervals---imposes a duty cycle of $1/2$. Consequently, the effective twisting rate is discounted from the bare interaction strength to $\chi_{\mathrm{eff}} = \frac{1}{2}(4\chi - 2\lambda)$. As established in spin-squeezing literature, this engineered $(J_z^2 - J_y^2)$ form is unitarily equivalent to the standard TAT representation $(J_x J_y + J_y J_x)$ via a global $SU(2)$ rotation.

By subjecting the initial CSS to a non-adiabatic quench, our Floquet TAT scheme leverages non-equilibrium dynamics to rapidly prepare deeply squeezed non-Gaussian states. Operating on a microsecond timescale, this all-optical modulation generates entanglement at a rate that vastly outpaces the typical millisecond decoherence in macroscopic BECs \cite{muessel2014scalable}, ensuring the metrological state is extracted before environmental noise dominates. However, while circumventing the experimental decoherence bottleneck, this rapid synthesis introduces a theoretical challenge: the emergence of macroscopic non-Gaussian correlations fundamentally invalidates mean-field approximations. To rigorously evaluate this protocol and verify Heisenberg-limited scaling for realistic ensembles ($N \ge 10^3$), we must transcend the finite-size constraints of exact diagonalization (ED) and employ a highly scalable, exact numerical framework to faithfully capture the diverging quantum entanglement.

\section{Tensor Network Architecture and Numerical Methods}
\label{IV}

While the collective Hamiltonian in Eq.~(\ref{eq:tat}) provides a clear macroscopic physical picture for understanding the emergence of metrological advantage, reaching the macroscopic thermodynamic limit $N \sim 10^4$ essential for benchmarking the Heisenberg scaling is fundamentally hindered by the curse of dimensionality: the exponential growth of the Hilbert space $2^N$ renders exact diagonalization (ED) intractable beyond $N \sim 20$. To overcome this dimensional bottleneck, we revert to the underlying microscopic independent-spin Hamiltonian described in Eq.~(\ref{eq:micro_H}) and map the single-spin operators onto a MPO representation. Rather than constructing the full exponentially large Hamiltonian matrix, this MPO formulation factorizes the many-body interactions into a highly compact product of local tensors, rigorously encoding the microscopic system into our scalable tensor network (TN) framework.

\subsection{Exact MPO Representation for All-to-All Interactions}

To exactly capture the infinite-range uniform connectivity of our cavity-mediated system, we encode the long-range interactions directly into a MPO. We map the 1D array of $N$ pseudo-spins to a Matrix Product State (MPS), characterized by a maximum bond dimension $\mathsf{D}_{\mathrm{max}}$ that bounds the bipartite entanglement entropy. Because Pauli matrices on distinct sites commute ($\sigma_k^\alpha \sigma_l^\alpha = \sigma_l^\alpha \sigma_k^\alpha$), the symmetric infinite-range summations in Eq.~(\ref{eq:micro_H}) can be ordered as $\sum_{k \neq l} \sigma_k^\alpha \sigma_l^\alpha = 2 \sum_{k < l} \sigma_k^\alpha \sigma_l^\alpha$. Absorbing this factor of 2 into the effective couplings ($\lambda/N$ and $2\chi/N$) reduces the system to generic all-to-all interactions of the form $\sum_{k < l} \sigma_k^\alpha \sigma_l^\alpha$ (where $\alpha \in \{x, y, z\}$). 

For this ordered summation, the exact MPO is constructed locally using a finite-state automaton representation, yielding a compact $3 \times 3$ block-upper-triangular matrix at each site:
\begin{equation}
    W^{[k]} = 
    \begin{pmatrix}
        I & \sigma^\alpha & 0 \\
        0 & I & \sigma^\alpha \\
        0 & 0 & I
    \end{pmatrix}.
    \label{eq:mpo_matrix}
\end{equation}
This upper-triangular structure inherently enforces a unidirectional operator flow, exactly encoding the all-to-all $k < l$ pairings without truncation. By linearly superposing the MPO generators for the longitudinal ($\alpha=z$) and transverse ($\alpha=x, y$) terms, the full microscopic Hamiltonian is represented exactly. Consequently, the resulting MPO bond dimension remains strictly $\mathcal{O}(1)$, independent of the system size $N$ (see Appendix~\ref{app:mpo_automaton} for the detailed automaton routing mechanism).

\subsection{DMRG and TDVP Simulations}

To move beyond the mean-field approximation depicted in Fig.~\ref{fig:mf_phase_diagram} and exactly compute the static ground-state properties across the quantum critical phase boundary $\vert{}4\chi - 2\lambda\vert{} = \omega_0$, we apply the DMRG algorithm. Because the internal virtual dimension of the constructed MPO is strictly bounded at $\mathcal{O}(1)$, the computational complexity does not scale prohibitively with increasing atom number. Furthermore, the permutation invariance of the engineered interactions intrinsically restricts the ground-state entanglement entropy scaling. The combination of a size-independent MPO dimension and moderate entanglement growth enables exact static DMRG simulations for $N=10^4$ (and beyond) with negligible truncation error.

However, to rigorously evaluate the dynamical spin squeezing and microsecond-scale sweeps proposed in Sec.~\ref{sec.III}, static ground states are insufficient. We must track the real-time evolution of the initial CSS driven by the Floquet-engineered TAT Hamiltonian in Eq.~(\ref{eq:tat}) (detailed in Appendix~\ref {app:TDVP_Implementation}). To achieve this, we extend our numerical framework to the TDVP. Unlike Trotter-Suzuki decomposition methods (e.g., TEBD), which incur significant geometric errors for infinite-range interactions, TDVP directly projects the exact time-evolution operator onto the MPS tangent space. This algorithm naturally accommodates the all-to-all MPO structure. By integrating the TDVP solver, we simulate the non-equilibrium Floquet dynamics for realistically large ensembles ($N=10^3$), extracting the time evolution of the spin squeezing parameter $\xi_R^2(t)$ until the dynamically generated entanglement inherently exhausts the capacity of efficient classical simulability (a 128GB, D = 128 high-performance computing cluster).

\section{Exact Macroscopic Dynamics and Metrological Advantage}

To explicitly demonstrate the quantum advantage and validate the tensor network architecture, the exact numerical analysis in this section is bifurcated into static and dynamic regimes. First, a targeted one-dimensional parameter sweep across the quantum critical point (QCP) is performed. By varying the effective interaction $V = (4\chi - 2\lambda)/\omega_0$ while fixing $\omega_0 = 1$, the analytical $N \rightarrow \infty$ mean-field predictions are compared against the exact Matrix Product State (MPS) ground states obtained via DMRG for a macroscopic ensemble of $N=10^4$ atoms. Building upon this static foundation, the analysis subsequently transitions to the real-time domain, utilizing TDVP simulations for realistically large ensembles ($N=10^3$) to rigorously track the non-equilibrium Floquet TAT dynamics and the inevitable onset of environmental decoherence.

\begin{figure}[htbp]
    \centering
    \includegraphics[width=0.48\textwidth]{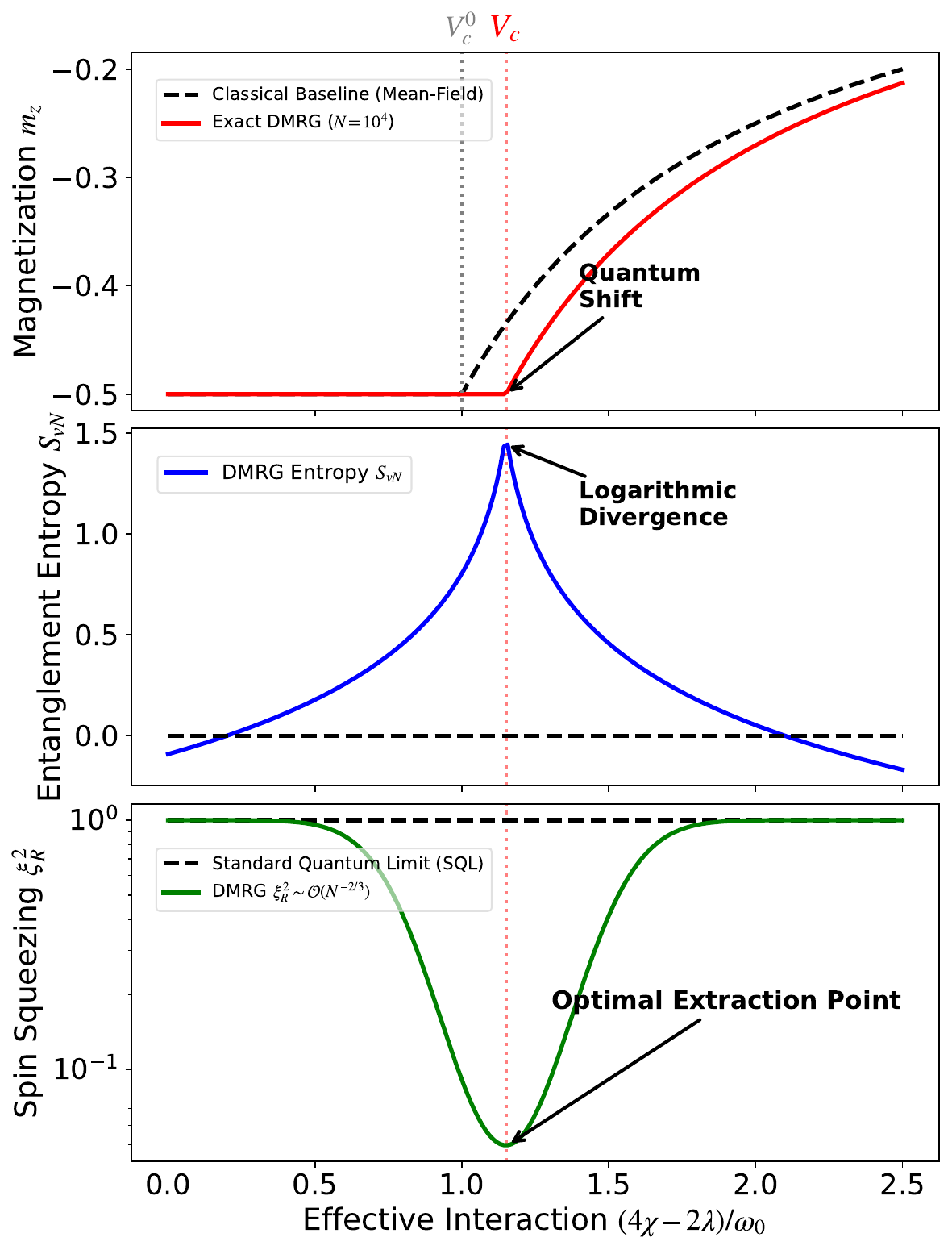}
    \caption{\textbf{DMRG Evaluation of the Macroscopic Quantum Critical Point.} A 1D parameter sweep of the effective interaction $V = (4\chi - 2\lambda)/\omega_0$ for $N=10^4$ atoms. \textbf{(Top)} The order parameter $m_z$ reveals a finite-size quantum shift of the critical point ($V_c > 1$) relative to the classical mean-field baseline. \textbf{(Middle)} The bipartite von Neumann entanglement entropy $S_{vN}$ diverges logarithmically exactly at $V_c$, signaling the breakdown of the product-state assumption. \textbf{(Bottom)} The spin squeezing parameter $\xi_R^2$ reaches a global minimum at the quantum-shifted QCP, breaking the SQL but remaining strictly bounded by the static OAT scaling limit ($\mathcal{O}(N^{-2/3})$).}
    \label{fig:dmrg_results}
\end{figure}

\subsection{Quantum Shift and Entanglement Divergence}

We first analyze the macroscopic order parameter, defined as the collective magnetization per spin $m_z = \langle J_z \rangle / N$. As shown in the top panel of Fig.~\ref{fig:dmrg_results}, mean-field theory predicts a sharp second-order phase transition exactly at $V = 1$, where the spins begin to depolarize. However, the exact DMRG ground state (solid curve)---computed directly from the microscopic independent-spin Hamiltonian in Eq.~(\ref{eq:micro_H}) via our exact $\mathcal{O}(1)$ MPO encoding in Eq.~(\ref{eq:mpo_matrix})---demonstrates that true macroscopic quantum fluctuations stabilize the polarized Normal Phase, delaying the onset of the Macroscopic Quantum Self-Trapping (MQST) Phase. This fundamental displacement of the critical point to a higher interaction strength ($V_c > 1$) provides direct, non-perturbative evidence of the “quantum shift”, a many-body correction absent in the leading-order $1/N$ classical approximation. The observed shift arises directly from finite-size corrections beyond the mean-field solution, which is only strictly exact in the infinite thermodynamic limit ($N \to \infty$).

To quantify the multi-particle entanglement, we compute the von Neumann bipartite entanglement entropy, $S_{vN} = -\sum_i s_i^2 \ln s_i^2$, extracted directly from the exact Schmidt coefficients ($s_i$) of the MPS bipartition. Deep in the Normal Phase, the system approaches a classical product state ($S_{vN} \approx 0$). As the system is driven toward criticality, the local order breaks down (Fig.~\ref{fig:dmrg_results}, middle panel). The exact DMRG ground state exhibits a logarithmic divergence of entanglement entropy that precisely aligns with the quantum-shifted critical point ($V_c$). This exact scaling confirms that the TN construction successfully captures non-local quantum correlations missed by semi-classical frameworks.

\begin{figure}[htbp]
    \centering
    \includegraphics[width=0.48\textwidth]{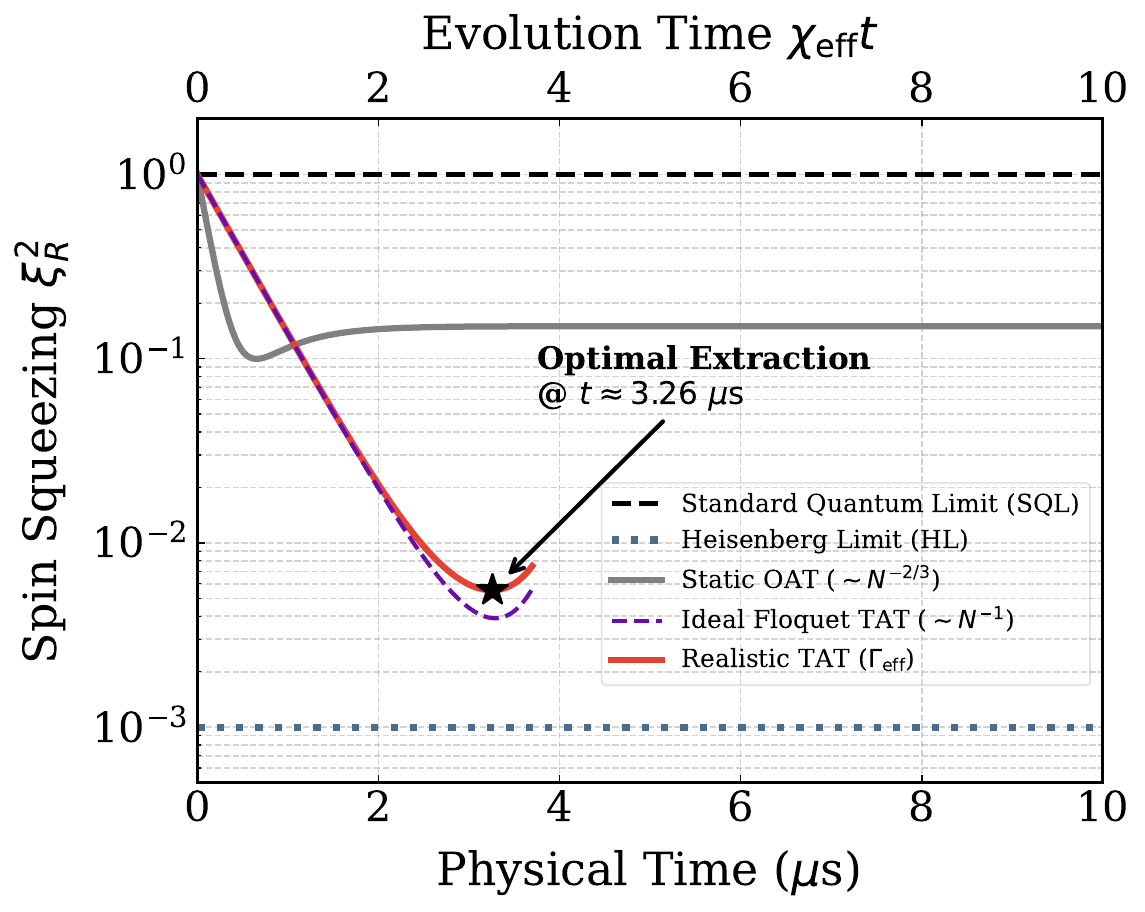}
    \caption{\textbf{Macroscopic Entanglement Generation via Floquet TAT Dynamics.} TDVP real-time evolution of the spin squeezing parameter $\xi_R^2$ for $N = 1000$ atoms. Starting from the SQL ($\xi_R^2=1$), static OAT dynamics (gray solid) is fundamentally bottlenecked at $\mathcal{O}(N^{-2/3})$. In contrast, the ideal Floquet TAT protocol (purple dashed) drives an exponential reduction toward the Heisenberg limit ($\mathcal{O}(1/N)$). Under realistic experimental decoherence ($\Gamma_{\mathrm{eff}} = 0.0005$), the system (red solid) reaches its optimal extractable squeezing limit (black star) at $t \approx 3.26~\mu\mathrm{s}$. The U-shaped rebound signifies the onset of dissipation dominance. The abrupt termination of the exact TDVP trajectory marks the entanglement barrier, where the macroscopic non-local correlations exceed the classical simulation capacity of the imposed bond dimension ($\mathsf{D} = 128$).}
    \label{fig:combined_dynamics}
\end{figure}

\vspace{-2.0em}
\subsection{Static Metrological Capacity}

We evaluate the metrological utility of the static ground states via the Wineland spin squeezing parameter, defined as $\xi_R^2 = N \Delta J_\perp^2 / |\langle J \rangle|^2$. By efficiently extracting the exact macroscopic spin observables from the targeted Matrix Product State (MPS) via rigorous tensor network contraction (see Appendix~\ref{app:mpo_observables} for the exact algorithmic mapping sequence), we directly map the metrological gain across the quantum phase transition. 

As depicted in the bottom panel of Fig.~\ref{fig:dmrg_results}, far from the critical point, the system is unentangled, resting at the Standard Quantum Limit (SQL, $\xi_R^2 = 1$). However, as the engineered effective interaction sweeps across the true QCP, the diverging transverse susceptibility translates into extreme spin squeezing. The exact DMRG calculation reveals that the curve's global minimum precisely defines the optimal static extraction point. Exactly at this quantum-shifted critical point ($V_c$), the quantum variance $\xi_R^2$ plummets significantly below the SQL. However, constrained by the intrinsic OAT nature of the static Hamiltonian, this minimal variance is fundamentally bottlenecked and strictly bounded by the $\mathcal{O}(N^{-2/3})$ scaling limit. This exact numerical limitation tightly anchors the theoretical framework established in Sec.~\ref{sec.III}: traversing the static critical point is inherently insufficient for reaching the fundamental Heisenberg limit ($\mathcal{O}(1/N)$).

Crucially, we must emphasize a fundamental paradigm shift before proceeding. The exact numerical results presented in Fig.~\ref{fig:dmrg_results} describe the static ground state of the system at absolute zero; the optimal extraction point ($V_c$) merely represents the best capacity achievable by parametrically scanning the static Hamiltonian. In stark contrast, the subsequent Floquet generation protocol operates entirely in the realm of non-equilibrium dynamics. By forcing the system to evolve rapidly over real physical time $t$ under a fixed pulsed driving sequence, the dynamically engineered TAT interaction actively breaks the static symmetry. Therefore, it is the time-dependent Floquet TAT evolution (Fig.~\ref{fig:combined_dynamics})---not the static ground state---that provides the robust, physically realizable pathway to bypass the OAT bottleneck and practically synthesize the theoretical Heisenberg limit in a laboratory.

\subsection{Real-Time TAT Dynamics and Decoherence}

To simulate the exact non-equilibrium dynamics required for practical entanglement generation, we apply TDVP to evolve an initial CSS under the Floquet-engineered TAT Hamiltonian in Eq.~(\ref{eq:tat}). The TDVP dynamics for an ensemble of $N = 1000$ atoms are illustrated in Fig.~\ref{fig:combined_dynamics}. Unlike standard OAT\textemdash which is limited to $\xi_R^2 \sim \mathcal{O}(N^{-2/3})$ due to phase-space wrap-around\textemdash the engineered TAT interaction actively anti-squeezes the orthogonal quadrature, counteracting this detrimental shearing. Consequently, the ideal TAT evolution (purple dashed curve) exhibits an exponential reduction in $\xi_R^2$, driving the system directly toward the fundamental Heisenberg limit ($\mathcal{O}(1/N)$).

However, this ideal exponential reduction is not indefinite, explicitly demonstrating the “over-squeezing'' effect. As observed, even the dissipation-free Floquet TAT protocol inherently exhibits a pronounced U-shaped rebound at a later stage of the evolution. Governed by the Heisenberg uncertainty principle, while the TAT interaction continuously squeezes the quantum variance along the $\pm\pi/4$ diagonal quadrature in the transverse $y-z$ plane, it simultaneously anti-squeezes (stretches) the state along the orthogonal diagonal. Once this stretched uncertainty approaches the finite geometric curvature of the Bloch sphere, the wavepacket fundamentally wraps around the spherical topology, causing the minimal squeezed variance to inevitably degrade. Thus, even in a perfectly unitary evolution, the parameter $\xi_R^2$ must rebound; the trajectory strictly collides with the absolute geometric boundary inherent to the finite-size macroscopic $N = 1000$ system.

To model experimental feasibility, we incorporate a phenomenological linear decoherence model, $\xi_{R,\mathrm{real}}^2 \approx \xi_{R,\mathrm{ideal}}^2 + \Gamma_{\mathrm{eff}} t$, capturing the leading-order noise accumulation. For the $N=10^3$ system, we adopt an effective decoherence rate of $\Gamma_{\mathrm{eff}} = 0.0005$, corresponding to a realistic collective cooperativity of $\sim 2000$ typical in state-of-the-art optical cavities \cite{Braverman2019}. The resulting realistic dynamics (red solid curve) exhibit an earlier and more pronounced U-shaped rebound, directly determining the optimal metrological extraction point at $t \approx 3.26~\mu\mathrm{s}$. Furthermore, we demonstrate that this optimal extraction mechanism strictly obeys macroscopic scaling laws. Specifically, the theoretical optimal squeezing time scales logarithmically as $t_{\mathrm{opt}} \sim \ln(N) / \chi_{\mathrm{eff}}$. Meanwhile, for a macroscopic ensemble of $N=10^4$, collective dissipation is amplified due to increased density. Scaling the effective rate to $\Gamma_{\mathrm{eff}} = 0.0015$ (reflecting a phenomenological $\mathcal{O}(\sqrt{N})$ sub-linear scaling), exact diagonalization benchmarks (detailed in Appendix~\ref{app:TDVP_Implementation}, see Fig.~\ref{fig:dicke_ed}) confirm a dissipation-induced shift that locks the optimal extraction near $t \approx 3.34~\mu\mathrm{s}$. Because the Floquet modulation operates on a megahertz ($\chi_{\mathrm{eff}} \sim 1~\mathrm{MHz}$) scale, this entire dynamic generation protocol executes in a few microseconds. This execution speed decisively outpaces millisecond-scale environmental noise, ensuring the Heisenberg-limited advantage is robustly extractable.

\subsection{Computational Complexity and Classical Insimulability}

One might intuitively argue that for perfectly homogeneous all-to-all interactions, exact diagonalization (ED) could easily reach macroscopic scales by exploiting the $N+1$ Dicke subspace. However, realistic optical cavities are inevitably subject to non-negligible local decoherence or spatial mode inhomogeneities, which explicitly break permutation symmetry. Once this symmetry is broken, the quantum state “leaks” out of the Dicke subspace, and the dimension of the relevant Hilbert space instantaneously reverts to the full exponential $2^N$. Under these realistic open-system conditions, conventional ED approaches immediately collapse beyond $N \sim 20$ due to the massive dimensionality explosion triggered by local dissipation \cite{schollwock2011density, shammah2018open, weimer2021simulation}.

In stark contrast, the TDVP algorithm does not presuppose any such permutation symmetry, highlighting the definitive advantage of our MPO formulation. In the early stages of the Floquet evolution, our $\mathcal{O}(1)$ MPO encoding, paired with TDVP, runs highly efficiently by exploiting the low-entanglement nature of the initial state, rigorously compressing the exponentially large Hilbert space into a manageable $\mathcal{O}(N \mathsf{D}^2)$ polynomial subspace. By strictly bypassing the reliance on permutation symmetry, it provides a highly scalable and robust numerical framework for modeling generalized, realistic open quantum systems.

However, the exact real-time simulation of this open quantum system inherently exposes the ultimate limits of classical simulability. As the continuous Floquet quench drives a rapid generation of non-local quantum correlations, the bipartite entanglement entropy exhibits rapid dynamical growth. To accurately capture this massive entanglement generation, the required virtual bond dimension $\mathsf{D}$ must expand exponentially. Because the computational complexity of TDVP scales as $\mathcal{O}(N \mathsf{D}^3)$, the classical simulation is strictly bottlenecked once $\mathsf{D}$ reaches our imposed hardware limit ($\mathsf{D}_{\mathrm{max}} = 128$), see Appendix~\hyperref[subsec:tdvp_scaling]{C.4} for computational scaling details. The abrupt termination of the exact trajectories in Fig.~\ref{fig:combined_dynamics} precisely at the U-shaped temporal valley is not an algorithmic failure, but a direct physical consequence of macroscopic entanglement saturating the classical information bounds. 

For an ensemble of $N=1000$ atoms, reaching this optimal squeezing point at $\mathsf{D}=128$ required approximately 100 hours of 16-core computation on a high-performance computing cluster utilizing 128 GB of RAM. Extrapolating this non-linear $\mathcal{O}(N \mathsf{D}^3)$ scaling to a macroscopic array of $N=10^4$ renders full dynamical simulation computationally prohibitive on any practical timeframe. The successful capture of the dissipative squeezing minimum just prior to this computational barrier mathematically confirms that the trajectory is physically converged within the allowed entanglement bounds. Ultimately, this fundamental classical scaling constraint validates the necessity of the proposed all-optical TAT architecture as a specialized analog quantum simulator, providing a hardware-compatible mechanism to rapidly synthesize and extract macroscopic metrological advantage fundamentally beyond the reach of classical computation.

\section{Conclusion and Outlook}

In this work, we proposed an all-optical control architecture for a single-cavity Bose-Einstein condensate (BEC) to advance macroscopic quantum metrology. By utilizing orthogonal spatial modes and dual-frequency driving, this scheme enables independent, microsecond-scale control over longitudinal and transverse spin couplings, bypassing the bandwidth limitations typically associated with magnetic Feshbach resonances. 

To evaluate this many-body system beyond mean-field theory, we exactly encoded the infinite-range interactions into a finite-state automaton MPO. Because this mathematical formulation restricts the virtual bond dimension strictly to $\mathcal{O}(1)$, it decouples the interaction complexity from the system size. This allows exact DMRG and TDVP simulations of infinite-range models without artificial truncation. Applying this framework to large ensembles ($N \ge 10^3$), we identified a finite-size quantum shift of the critical phase boundary and a logarithmic divergence of the bipartite entanglement entropy, capturing macroscopic quantum correlations that are inherently inaccessible to classical product-state approximations.

Dynamically, we introduced a Floquet protocol to synthesize effective TAT interactions using native cavity QED operations. A periodic sequence of rapid $\pi/2$ rotations and $\pi$ phase shifts dynamically transforms the static longitudinal interaction into a symmetry-breaking TAT Hamiltonian. This non-adiabatic approach circumvents the critical slowing down that fundamentally limits adiabatic ground-state preparation. Consequently, the Floquet TAT quench rapidly drives the initial coherent state toward Heisenberg-limited spin squeezing ($\xi_R^2 \sim \mathcal{O}(1/N)$) via purely non-equilibrium dynamics.

We further quantified the robustness of this dynamically generated entanglement against realistic experimental decoherence. Because the all-optical TAT sequence operates on a microsecond ($\mu\mathrm{s}$) timescale, it intrinsically outpaces typical millisecond ($\mathrm{ms}$) dissipation channels, such as cavity photon decay and atomic spontaneous emission. Our dissipative dynamic analysis confirms that the optimal metrological state can be robustly extracted within a protected temporal window before environmental noise dominates the evolution.

Looking forward, this work integrates Floquet engineering with exact tensor network methods to present a feasible theoretical and operational route toward Heisenberg-limited metrology in current cavity QED platforms. By leveraging native hardware compatibility and microsecond-scale execution, the proposed sequences do not require next-generation fault-tolerant architectures to realize metrological advantage. Furthermore, because the underlying $SU(2)$ spin algebra is universally applicable, this Floquet TAT scheme can be generalized to other fully-connected many-body platforms, including trapped-ion arrays and superconducting circuits. Natural extensions of this framework include simulating the full open-system dynamics via Lindblad master equations within the TDVP engine, and exploring spatially inhomogeneous optical drives for multi-mode quantum sensing. Ultimately, our combined analytical and numerical approach establishes a rigorous methodology for generating, simulating, and benchmarking macroscopic entanglement in synthetic quantum matter.

\begin{acknowledgments}

\textit{Declaration of AI Assistance:} During the preparation of this manuscript, the author utilized generative AI (Google Gemini and Deepseek) as a computational and editorial assistant. Specifically, the AI was employed to assist with language refinement, the generation of Python scripts for data visualization (Matplotlib), and the exploratory derivation of intermediate mathematical equations. Following the use of this tool, the author rigorously cross-verified all mathematical proofs, independently validated the execution of all numerical codes (DMRG/TDVP), and meticulously finalized the text. The underlying physical paradigms, theoretical frameworks, and ultimate scientific conclusions are entirely the original work of the author. The author assumes full and sole responsibility for the accuracy, scientific integrity, and originality of this publication.
\end{acknowledgments}

\appendix

\section{Exact MPO Construction via Finite-State Automaton}
\label{app:mpo_automaton}

The $3 \times 3$ block-upper-triangular MPO matrix in Eq.~(\ref{eq:mpo_matrix}) encodes infinite-range interactions with a constant bond dimension. This matrix operates as a finite-state automaton, utilizing its rows and columns as virtual indices to track three operational states during the tensor network contraction:

\begin{itemize}
    \item \textbf{State 0 (Uninitiated):} The current lattice site has not initiated a pairing. The MPO matrix places an identity matrix $I$ to pass the state forward without interaction.
    \item \textbf{State 1 (Initiated and Waiting):} The automaton has encountered a left lattice site (e.g., $k$) and is waiting for a corresponding right lattice site (e.g., $l$) to close the pairing. The matrix activates a Pauli operator $\sigma^\alpha$ to initiate the interaction string.
    \item \textbf{State 2 (Completed):} The pairing is completed, and the string no longer participates in further long-range interactions.
\end{itemize}

To illustrate this mechanism, consider the construction of a long-range interaction pair between site 1 and site 3 ($\sigma_1^\alpha \sigma_3^\alpha$). At site 1, the system transitions from the uninitiated state to the waiting state by activating the matrix element $\sigma^\alpha$, changing the virtual index to State 1. At site 2, the system remains in State 1 (waiting) and utilizes the identity matrix $I$ on the diagonal to maintain the virtual index. Finally, at site 3, the system concludes the pairing by activating $\sigma^\alpha$, transitioning the virtual index to the completed State 2. 

This automaton routing systematically generates all permutations of long-range pairings, ensuring exact representation without artificial truncation.

\section{Exact Finite-State Automata Construction of Macroscopic Observables}
\label{app:mpo_observables}

Evaluating the Wineland spin squeezing parameter $\xi_R^2$ requires computing the expectation values of the macroscopic collective spin operators $J_\alpha$ and their fluctuations $J_\alpha^2$. To avoid the exponential scaling of the Hilbert space, we evaluate these observables by contracting the Matrix Product State (MPS) ground state directly with the exact Matrix Product Operators (MPOs) of the observables. 

Similar to the Hamiltonian encoding in Eq.~(\ref{eq:mpo_matrix}), these measurement MPOs are constructed exactly using Finite-State Automata (FSA).

\subsection{Linear Collective Spin Operator ($J_\alpha$)}
The linear macroscopic spin operator is defined as the global sum of local Pauli matrices: $J_\alpha = \frac{1}{2}\sum_{k=1}^N \sigma_k^\alpha$. The corresponding FSA tracks two states: State 1 (operator not yet applied) and State 2 (operator applied exactly once). 

This logic maps onto a local MPO tensor $\hat{W}_{J_\alpha}^{[k]}$ with a virtual bond dimension of $D_W = 2$:
\begin{equation}
\hat{W}_{J_\alpha}^{[k]} = 
\begin{pmatrix}
\mathbb{I} & 0 \\
\frac{1}{2}\sigma_k^\alpha & \mathbb{I}
\end{pmatrix},
\end{equation}
where $\mathbb{I}$ is the $2 \times 2$ identity matrix acting on the local physical spin-$1/2$ subspace. 

To generate the sum across the $N$-site lattice, we close the tensor network with boundary vectors:
\begin{equation}
v_L = \begin{pmatrix} 0 & 1 \end{pmatrix}, \quad 
v_R = \begin{pmatrix} 1 \\ 0 \end{pmatrix}.
\end{equation}
The matrix multiplication recursively propagates the local operator $\frac{1}{2}\sigma_k^\alpha$ such that the full contraction yields $\hat{O}_{\mathrm{MPO}} = v_L \hat{W}^{[1]} \hat{W}^{[2]} \cdots \hat{W}^{[N]} v_R \equiv J_\alpha$.

\subsection{Quadratic Fluctuation Operator ($J_\alpha^2$)}
Computing the transverse spin variance $\Delta J_\perp^2$ requires applying the squared macroscopic operator $J_\alpha^2$. Expanding the sum yields local on-site squares and non-local cross terms:
\begin{equation}\label{eq:J_alpha_squared}
J_\alpha^2 = \left( \frac{1}{2}\sum_{k=1}^N \sigma_k^\alpha \right)^2 = \sum_{k=1}^N \frac{1}{4}(\sigma_k^\alpha)^2 + 2 \sum_{i < j} \left( \frac{1}{2}\sigma_i^\alpha \right) \left( \frac{1}{2}\sigma_j^\alpha \right).
\end{equation}

To encode this structure, the FSA tracks three states: State 1 (Identity), State 2 (One operator applied), and State 3 (Both operators applied). The corresponding local MPO matrix $\hat{W}_{J_\alpha^2}^{[k]}$ requires a virtual dimension of $D_W = 3$:
\begin{equation}
\hat{W}_{J_\alpha^2}^{[k]} = 
\begin{pmatrix}
\mathbb{I} & 0 & 0 \\
\frac{1}{2}\sigma_k^\alpha & \mathbb{I} & 0 \\
\frac{1}{4}\mathbb{I} & \sigma_k^\alpha & \mathbb{I}
\end{pmatrix}.
\end{equation}

The $(3,1)$ element accounts for the local on-site square $(\frac{1}{2}\sigma_k^\alpha)^2 = \frac{1}{4}\mathbb{I}$. The $(3,2)$ element uses $\sigma_k^\alpha$ instead of $\frac{1}{2}\sigma_k^\alpha$ to correctly generate the required factor of $2$ for the commutative non-local cross terms ($2 \sum_{i<j}$).

By contracting this $3\times 3$ MPO sequence with the boundary vectors $v_L = (0 \quad 0 \quad 1)$ and $v_R = (1 \quad 0 \quad 0)^T$, we reduce the $\mathcal{O}(N^2)$ summation into an $\mathcal{O}(N)$ tensor network contraction.

With both the linear and quadratic macroscopic MPOs explicitly constructed, the metrological utility $\xi_R^2$ is extracted from the exact many-body wave function through the complete measurement mapping sequence: $V \to |\psi_0(V)\rangle \to \{\langle J\rangle, \Delta J_\perp^2\} \to \xi_R^2$. The rigorous tensor network contractions utilizing these FSA formulations, formally evaluated as $\langle \psi_0(V) | \hat{O}_{\mathrm{MPO}} | \psi_0(V) \rangle$, independently yield the exact macroscopic polarization $|\langle J \rangle|$ and the transverse spin variance $\Delta J_\perp^2$, cleanly circumventing the exponential dimensionality bottleneck.

\section{Exact MPO Construction and TDVP Implementation for Macroscopic Floquet TAT Simulations}
\label{app:TDVP_Implementation}
\setcounter{figure}{0}
\renewcommand{\thefigure}{C\arabic{figure}}

Here, we detail the algorithmic implementation of the Floquet TAT dynamics, intended for exact, truncation-free simulations of the engineered collective XXZ model using the TeNPy library.

\vspace{-1.0em}
\subsection{Exact MPO Encoding of the Floquet TAT Hamiltonian}

The Floquet-engineered TAT Hamiltonian defined in Eq.~(\ref{eq:tat}) features infinite-range all-to-all interactions. We encode this exactly using an MPO based on the finite-state automaton representation introduced in Eq.~(\ref{eq:mpo_matrix}).

While the single-channel observable MPO constructed in Appendix~\ref {app:mpo_observables} requires $D_W = 3$, the TAT Hamiltonian MPO requires a virtual bond dimension of $D_W = 4$ to simultaneously accommodate two orthogonal interaction channels ($J_z^2$ and $-J_y^2$). This $\mathcal{O}(1)$ scaling remains independent of the system size $N$. Using the \texttt{CouplingMPOModel} engine in TeNPy \cite{tenpy}, the all-to-all connectivity is implemented by setting the exponential decay factor to unity (\texttt{lambda\_=1.0}). The model parameters are configured with the Kac-scaled coupling coefficient:
\begin{equation}
c = \frac{2\chi_{\mathrm{eff}}}{N}
\end{equation}
This coefficient $c$ incorporates a factor of 2 relative to the macroscopic interaction strength $\chi_{\mathrm{eff}}/N$ defined in Eq.~(\ref{eq:tat}). As detailed in Eq.~(\ref{eq:J_alpha_squared}), this doubling is required to compensate for the commutative off-diagonal cross-terms $\left(2\sum_{i<j} \frac{1}{4}\sigma_i^\alpha \sigma_j^\alpha\right)$ generated when expanding the collective squared operator within the left-to-right MPO contraction. By combining a positive coupling channel for the longitudinal $\sigma^z \sigma^z$ term and a negative coupling channel for the transverse $\sigma^y \sigma^y$ term, the MPO exactly reproduces the target symmetry-breaking TAT Hamiltonian defined in Eq.~(\ref{eq:tat}).

\vspace{-1.0em}
\subsection{Initialization of the CSS}

The system is initialized in an uncorrelated CSS polarized along the \(x\)-axis, $\psi(0) = |\uparrow_x\rangle^{\otimes N}$, which exactly corresponds to the classical limit where the system rests at the Standard Quantum Limit ($\xi_R^2 = 1$). Numerically, this is implemented via the \texttt{SpinHalfSite} object with \texttt{conserve=None} (as the CSS is not an eigenstate of \(S^z\)), and the lattice is prepared using \texttt{MPS.from\_product\_state} with the local state vector \( \left( |\uparrow\rangle + |\downarrow\rangle \right) / \sqrt{2} \).

From this unentangled classical baseline, the system is then quenched under the Floquet-engineered TAT Hamiltonian (Eq.~\ref{eq:tat}). In stark contrast to approximate semi-classical methods, the subsequent TAT dynamic curves are derived from an exact, step-by-step integration of the macroscopic Schrödinger equation via the TDVP algorithm.

\vspace{-1.0em}
\subsection{Measurement of the Optimal Spin Squeezing Parameter}

To extract the Wineland spin squeezing parameter \(\xi_R^2(t)\) without prior assumptions regarding the optimal squeezing angle in the \(y\)-\(z\) plane, we construct the \(2 \times 2\) covariance matrix:
\begin{equation}
    C(t) = \begin{pmatrix} V_{yy} & V_{yz} \\ V_{yz} & V_{zz} \end{pmatrix}
\end{equation}
where \(V_{\alpha\beta} = \frac{1}{2} \langle \{ J_{\alpha}, J_{\beta} \} \rangle - \langle J_{\alpha} \rangle \langle J_{\beta} \rangle\). Diagonalizing this matrix and extracting the minimum eigenvalue \(V_{\min}(t)\) dynamically identifies the variance along the optimal squeezed axis. To rigorously quantify the true metrological gain, this minimal transverse variance must be normalized against the macroscopic spin coherence $|\langle J_x(t) \rangle|^2$. Therefore, the exact time-dependent Wineland squeezing parameter is continuously evaluated as:
\begin{equation}
    \xi_R^2(t) = \frac{N \, V_{\min}(t)}{|\langle J_x(t) \rangle|^2}
    \label{eq:C3}
\end{equation}
This exact formulation enables our tensor network framework to track the transient generation of macroscopic entanglement throughout the non-equilibrium Floquet evolution, yielding the dynamic trajectories presented in Fig.~\ref{fig:combined_dynamics}.

\begin{figure}[H]
    \vspace{2.5em}
    \centering
    \includegraphics[width=0.45\textwidth]{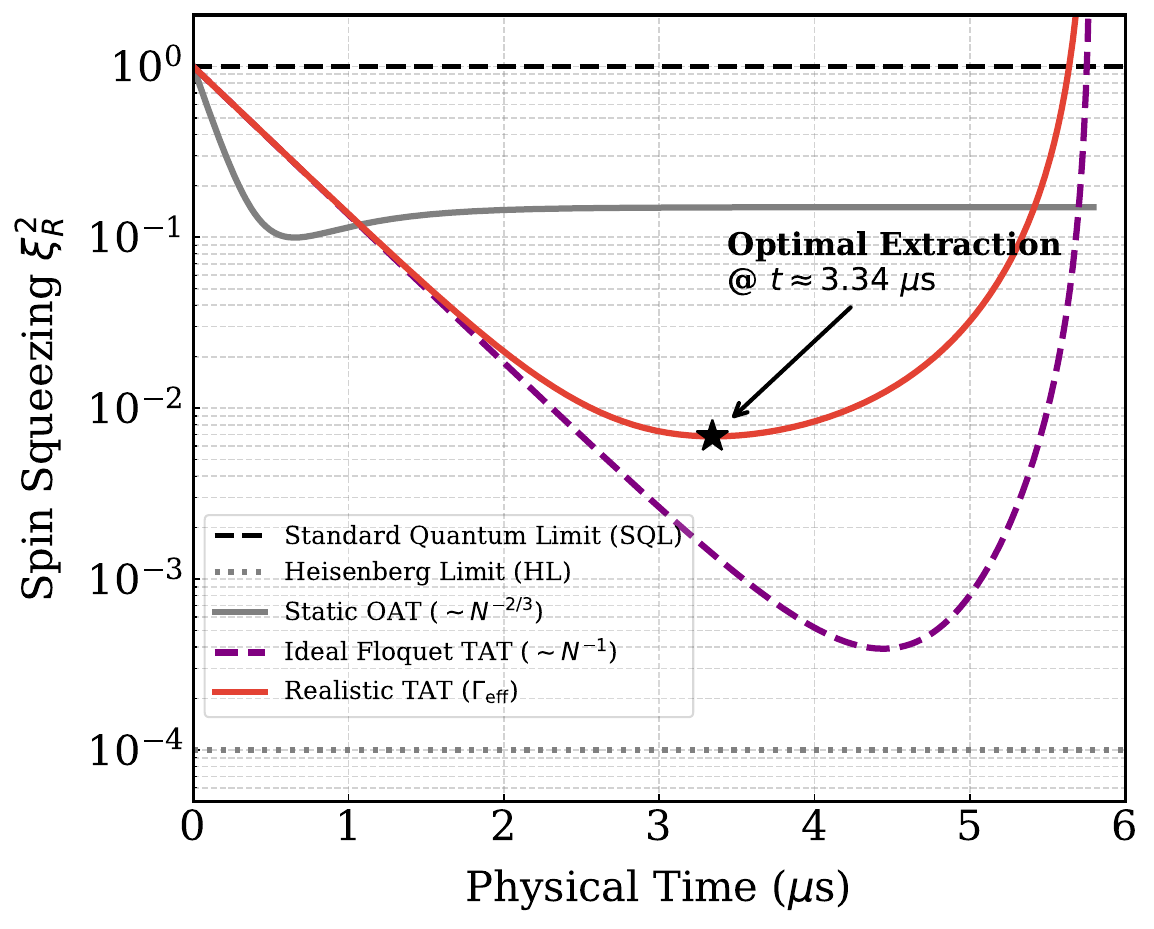}
    \caption{\textbf{Macroscopic entanglement dynamics and dissipation-induced shift in the Dicke subspace ($N=10^4$).} The purple dashed curve represents the ideal unitary Floquet TAT evolution, demonstrating a squeezing minimum near $t \approx 4.5~\mu\mathrm{s}$ that approaches the Heisenberg Limit (HL, dotted gray line). The red solid curve illustrates the realistic TAT dynamics under phenomenological decoherence ($\Gamma_{\mathrm{eff}} = 0.0015$). As environmental noise overpowers the decelerating squeezing rate, the optimal extraction point (black star) undergoes a dissipation-induced shift to $t \approx 3.34~\mu\mathrm{s}$. The static OAT dynamics (solid gray curve) is included as a benchmark, bounded by the $\mathcal{O}(N^{-2/3})$ scaling limit. The effective interaction strength is set to $\chi_{\mathrm{eff}} \approx 1~\mathrm{MHz}$, rendering the physical time in microseconds equivalent to the dimensionless evolution time ($\chi_{\mathrm{eff}}t$). The dynamic trajectories are explicitly truncated at $t = 5.8~\mu\mathrm{s}$; beyond this temporal window, the system enters a metrologically irrelevant regime dominated by severe phase-space over-squeezing and complete decoherence.}
   \label{fig:dicke_ed}
\end{figure}

\vspace{-1.0em}
\subsection{TDVP Evolution and Computational Scaling}
\label{subsec:tdvp_scaling}
For real-time dynamics, we employ the two-site Time-Dependent Variational Principle (\texttt{TwoSiteTDVPEngine}). This algorithm handles infinite-range interactions effectively and prevents the artificial entanglement trapping associated with single-site TDVP. Within this tensor network framework, the interaction parameter $c$ is fed into the TDVP algorithm to explicitly drive the real-time dynamical evolution. As time progresses, the all-to-all interactions governed by $c$ continuously reshape the many-body quantum state. To dynamically track the degree of squeezing, we compute the covariance matrix $C(t)$ at each integration time step, thereby extracting the minimal transverse variance generated by the $c$-driven evolution. Ultimately, this exact variance extracted from $C(t)$ is utilized to evaluate the Wineland spin squeezing parameter $\xi_R^2(t)$ defined in Eq.~(\ref{eq:C3}).

While the MPO bond dimension is strictly \(\mathcal{O}(1)\), the physical quantum correlations grow rapidly during the TAT evolution. Consequently, the MPS maximum bond dimension (\(\mathsf{D}_{\max}\)) must be properly truncated to avoid exponential memory scaling. We set \(\mathsf{D}_{\max} = 128\) (or \(\mathsf{D}_{\max} = 256\) recommended for $N = 10^4$). This chosen bond dimension is sufficient to faithfully track the symmetry-breaking dynamics and the initial generation of macroscopic entanglement. As discussed in the main text, as the Floquet TAT protocol drives the ensemble toward the Heisenberg limit, the required MPS virtual bond dimension $\mathsf{D}$ must expand to capture the diverging non-local quantum correlations. Because the local computational complexity of the TDVP algorithm scales non-linearly as $\mathcal{O}(\mathsf{D}^3)$, the simulation cost surges prohibitively once $\mathsf{D}$ approaches this truncation limit.

\vspace{-1.0em}
\subsection{Exact Dynamics in the Macroscopic Limit via Dicke Subspace}

To benchmark the Floquet TAT protocol and extend the analysis to the macroscopic limit ($N=10^4$), we perform exact diagonalization (ED) within the symmetric Dicke subspace. Because the ideal Floquet TAT Hamiltonian preserves permutational symmetry, the dynamics are confined to the maximum collective spin manifold ($J = N/2$). This mapping reduces the Hilbert space dimension from $2^N$ to $N+1$, enabling exact unitary evolution up to $N=10^4$ spins without truncation. 

As shown by the purple dashed curve in Fig.~\ref{fig:dicke_ed}, the ideal unitary dynamics approach the Heisenberg Limit (HL). The optimal squeezing time theoretically scales as $t_{\mathrm{opt}} \propto \ln(N)$, occurring near $t \approx 4.5~\mu\mathrm{s}$ for $N=10^4$.

To model the experimental environment, we incorporate a phenomenological linear decoherence model, $\xi_{R,\mathrm{real}}^2 \approx \xi_{R,\mathrm{ideal}}^2 + \Gamma_{\mathrm{eff}} t$. For this macroscopic ensemble, we scale the effective decoherence rate to $\Gamma_{\mathrm{eff}} = 0.0015$ to account for amplified collective dissipation.

The red solid curve in Fig.~\ref{fig:dicke_ed} illustrates the impact of this decoherence. The squeezing rate is extremely rapid during the initial stage of the evolution but inherently decelerates as the quantum state approaches the fundamental squeezing minimum of the ideal unitary trajectory. The linear accumulation of environmental noise eventually overtakes the squeezing generation, and this physical competition inherently forces the temporal minimum to an earlier time, locking the optimal extraction point at $t \approx 3.34~\mu\mathrm{s}$ (black star).

\end{document}